\documentclass[aip,apl,amsmath,amssymb,reprint,superscriptaddress]{revtex4-1}%

\usepackage{graphicx}
\usepackage{dcolumn}
\usepackage{bm}

\begin{document}

\preprint{AIP/123-QED}

\title[Sample title]{ Tunable degree of localization in random lasers with controlled interaction}

\author{Marco Leonetti}
\affiliation{ISC-CNR, UOS Sapienza, P. A. Moro 2, 00185 - Roma, Italy}
\affiliation{Instituto de Ciencia de Materiales de Madrid (CSIC)
 and Unidad Asociada CSIC-UVigo, Cantoblanco 28049 Madrid Espa\~{n}a.}

\homepage[]{www.luxrerum.org}\email[]{marco.leonetti@icmm.csic.es}

\author{Claudio Conti}
\affiliation{Dep. Physics University Sapienza, P.le Aldo Moro 5, I-00185, Roma Italy}
\affiliation{ISC-CNR, UOS Sapienza, P. A. Moro 2, 00185 - Roma, Italy}

\author{Cefe L\'opez}%
\affiliation{Instituto de Ciencia de Materiales de Madrid (CSIC)
 and Unidad Asociada CSIC-UVigo, Cantoblanco 28049 Madrid
Espa\~{n}a.}


\date{\today}

\begin{abstract}
We show that the degree of localization for the modes of a random laser (RL) is affected by the inter mode interaction that is controlled by shaping the spot of the pump laser. By experimentally investigating the spatial properties of the lasing emission  we infer that strongly localized modes are activated in the low interacting regime while in the strongly interacting one extended modes are found lasing. Thus we demonstrate that the degree o localization may be finely tuned at the micrometer level.
\end{abstract}

\maketitle

RL are among the most complex  systems in photonics, encompassing structural disorder and nonlinearity,\cite{Conti2011} and ranging from micron sized optical cavities \cite{Cao_Confinement} to kilometer-long fibers \cite{Turitsyn2010}. Attention on this systems has been constantly growing as the number of potential applications, ranging from object coding \cite{springerlink} to speckle-free illumination \cite{Redding2012}.

\begin{figure}
\includegraphics[width=\columnwidth]{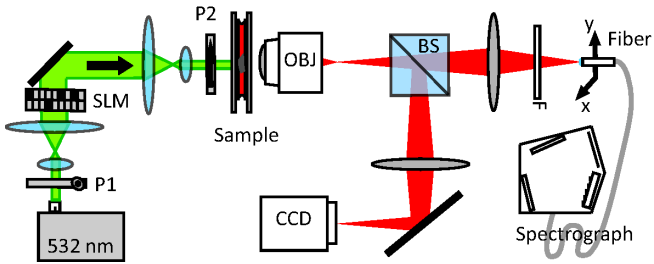}
\caption{(Color Online) Experimental setup. The pulsed laser (532 nm, repetition frequency 10 Hz and fluence $0.1 nJ/ \mu m^2$), whose spot  is shaped by a SLM in amplitude configuration (by using the two crossed polarizers P1 and P2), pumps a single titanium dioxide cluster (diameter between 5 and 12 $\mu$m). The RL emission is collected by a microscope objective (OBJ) to be imaged by using a beamsplitter (BS) in two different image planes. In one of them lies a fiber controlled by translators with nanometric resolution that allows the measurement of the spatio-spectral map. In this way it is possible to scan a magnified (50$\times$) image of the sample and measure the spectra emitted from a single point. The fibre core (50 $\mu$m in diameter) collects spectra originating in an area of 1$\mu$m of diameter of the sample. The other light path allows imaging the sample on a CCD.
\label{setup}}
\end{figure}

First-principle time domain simulations show that modes of a RL arise from electromagnetic states\cite{Conti08,Andreasen10,Lagendijk_2007_extent}, that may appear in localized or extended fashion.  Several experiments confirmed this view \cite{Fallert_coexistence_nature,Cao_Confinement,PhysRevB.59.15107}, and tried to address the connection between the structure \cite{Labonte:12} pumping condition, or gain and the degree of localization \cite{PhysRevA.81.043818, Yamilov:05, PSSB:PSSB200983268}. The presence of many modes may give rise to unique phenomena: in a linear system extended necklace states spread over the sample via multiple (localized) resonances \cite{PhysRevLett.94.113903} while in a system  with gain the inter-mode coupling affects the whole spectrum generating mode repulsion \cite{PhysRevB.67.161101} directly connected to the nonlinear interaction \cite{Türeci02052008}.

On the other hand it has been recently demonstrated that the inter-mode coupling plays a critical role in the onset of two fundamentally different RL regimes, distinguished by the shape of the emission: a ``resonant feedback random laser'' (RFRL) \cite{Cao_firstresonant}, which appears as a set of sharp peaks oscillating independently at fixed spectral positions, and the ``intensity feedback random laser'' (IFRL) \cite{Lawandy_Nature}, characterized by a smooth single line narrowed spectrum. By using a tailored spatial shape of the pump area \cite{Leonetti2011,PhysRevA.85.043841}, a switching between the two can be achieved. In fact a RFRL is observed when activating a set of weakly interacting resonances while IFRL is produced under strong interaction that leads to a mode-locked synchronized regime.

\begin{figure}
\includegraphics[width=\columnwidth]{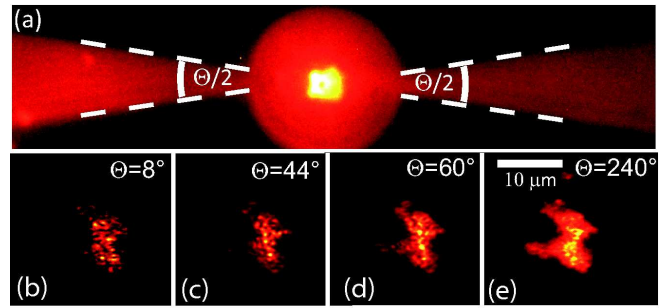}
\caption{(Color Online) a) A cluster and surrounding pumped area for $\Theta=36^\circ$ . b),c),d), and e):  Images of the random lasing intensity emitted from the cluster for different values of $\Theta$ for the cluster C1. \label{fig 5}}
\end{figure}

In this Letter, we present the experimental evidence that the mode extent is affected by the inter mode interaction. We demonstrate that the degree of localization may be finely tuned by controlling the pump shape:  when the number of modes activated in a single isolated disordered structure is increased, the extent of a single mode is also augmented.  Our set up comprises a spatial light modulator (SLM) that is imaged as a pumped area in a rhodamine solution pool where a single cluster of titanium dioxide nanoparticles is embedded. The illuminated area is composed by a circular spot centered on the cluster that prepares the system barely below the lasing threshold and  by two wedges of angular aperture $\Theta$ that pump the rhodamine bath surrounding the cluster, so that amplified spontaneous emission of the dye pumps the region in a directional and controlled manner (see Fig. \ref{setup} for a scheme of the setup and Fig.  \ref{fig 5}a for an image of the pump shape). The control on the degree of mode interaction is exerted by varying the angular span of the illuminated wedges thus increasing the number of excited cluster modes which, through their mutual interaction, mediate the involvement of more modes in the lasing action.

A recently developed approach (details in \cite{PhysRevA.85.043841}) allows to fix the amount of lasing modes brought over the threshold  by controlling  the $\Theta$  parameter. We show in the following that the spatial shape of the intensity on the sample at a fixed wavelength is strongly affected by the pumping condition and by the strength of the interaction. In figure \ref{fig 5} b-e an image of the cluster (sample C1) is shown for various values of $\Theta$. While at low $\Theta$ the image pattern displays hotsposts embedded in low intensity regions, when $\Theta$ is larger the intensity shows a smoother profile where no dark areas are evident.

\begin{figure}
\includegraphics[width=\columnwidth]{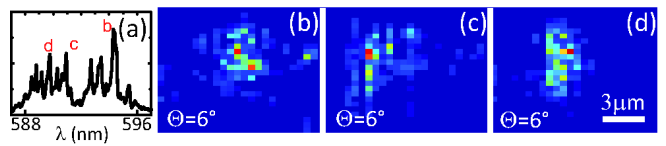}
\caption{(Color Online) (a) Spectrum from the cluster C3 obtained by pumping with $\Theta = 6 ^\circ$;  (b,c,d) spatial distribution of intensity for the three modes indicated in (a)\label{fig 1} }
\end{figure}

In a RFRL, modes have sub-nanometer FWHM and lie in a limited spectral range (see Fig. \ref{fig 1}a), thus  simple image analysis does not allow to obtain quantitative information about the spatial extent of every single mode; to this aim both spatial  and spectral resolution is needed \cite{Lagendijk_2010_modestructure}. This is achieved by a suitably designed experimental setup: an image of the sample is formed trough a 50$\times$ magnification optics (microscopy objective and eyepiece) in an plane that is scanned by a motorized fiber connected to a spectrograph. The result of a scan session is a complete spatio-spectral map.

\begin{figure}
\includegraphics[width=\columnwidth]{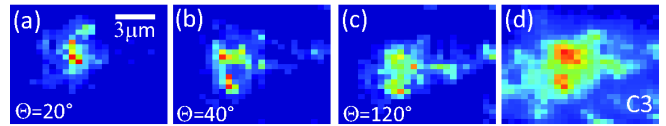}
\caption{ (Color Online) a , b, and c  represent the spatial intensity distribution for the most intense mode of cluster C3 when pumped with $\Theta = 20^\circ$, $\Theta = 40^\circ$,  and $\Theta = 120 ^\circ$ respectively, while  panel d, reports the spatial distribution of the intensity (all wavelengths summed) below lasing threshold providing the shape of the cluster.\label{fig 2} }
\end{figure}

The spatial extent of a mode resonant at $\lambda$ may be obtained by selecting the intensity emitted at the corresponding wavelength. Examples of the map for different modes (frequencies) are given in  Fig.\ref{fig 1}. Panel a shows the complete spectrum emitted by (whole \footnote[1]{Obtained summing all the contribution at different points}) cluster C3 (whose shape is reported in figure \ref{fig 2} d) for $\Theta=6^\circ$ , panels \ref{fig 1}b,c,d show the spatial distribution of the intensity for the peaks labeled in the spectrum. Each pixel in the images corresponds to a $1\times1 \mu m$ area. The three most intense modes dwell in different areas of the cluster. Figures \ref{fig 2}a ,b and c show the spatial extent of the most intense mode of the cluster C3 for $\Theta=20^\circ$, $\Theta=40^\circ$ and $\Theta=120^\circ$. The physical shape of the cluster can be seen in panel \ref{fig 2}d, which shows the image of the scattered light obtained by pumping at fluences  well below the lasing threshold and summing contribution at all wavelengths. Previous work showed that the $\Theta$ parameter governs the RL regime allowing the transition from RFRL to a IFRL\cite{Leonetti2011}. Fig \ref{fig 2} shows that this parameter also affects the spatial properties of single modes.

The degree of localization of a single mode at wavelength $\lambda$ with spatial profile of intensity $I_\lambda(x,y)$, is given by the inverse participation ratio $P$($\mu$m$^{-2}$)

\begin{eqnarray}
P_\lambda(\Theta)\equiv\frac{\int I_\lambda(x,y)^2}{(\int I_\lambda(x,y))^2} dx dy
\end{eqnarray}
and by the localization length $\Omega_\lambda$($\mu$m) \cite{Gentilini_09}
\begin{eqnarray}
\Omega_\lambda(\Theta)=1/\sqrt{P_\lambda}\text{.}
\end{eqnarray}

\begin{figure}
\includegraphics[width=\columnwidth]{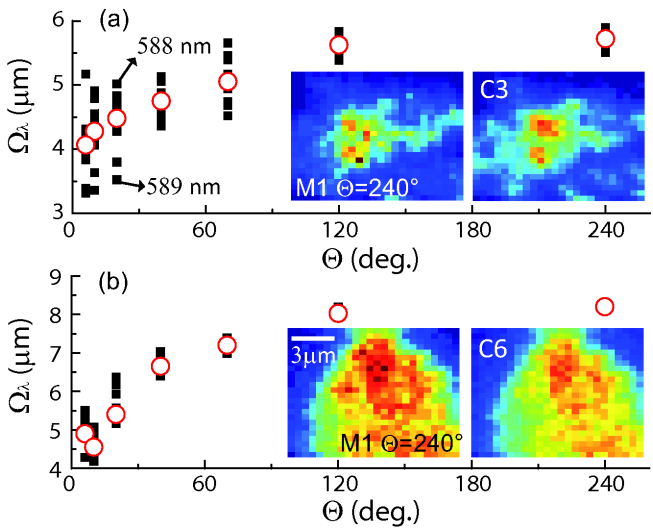}
\caption{(Color Online) Localization length $\Omega$ versus angular spread of pump $\Theta$ for samples C3 (a) and C6 (b). Left insets show the shape of the most intense mode (M1) for $\Theta=240^\circ$, while right insets show the samples shape.
\label{fig 3}}
\end{figure}

Fig.\ref{fig 3} shows $\Omega$ as a function of $\Theta$ for two different clusters, C3 (Fig.\ref{fig 3}a) and C6 (Fig.\ref{fig 3}b). Black squares refer to the ten most intense modes, their average for each angular span being represented by the open circles. For C3, $\Omega_\lambda$ grows from ca. 4 to 6$\mu$m (50\% increment of the mode extension), while  for C6 it grows from ca. 4.5 to 8$\mu$m (about 80\% increment in the mode extension). In the latter case, the larger $\Omega$ is ascribed to the larger size of C6, in fact when $\Theta$ is sufficiently large, modes spread over the whole sample and their shape resembles that of the cluster (see insets of figure \ref{fig 3}). These results are valid not only for all the $10$ most intense modes for C3 and C6, but also for all the modes observed in four other clusters with comparable spatial dimension (not reported).

It is well known that factors like gain \cite{PSSB:PSSB200983268} or losses \cite{Yamilov:05} may influence the spatial shape of modes. It is clear that in the present case none of this features may provide an explanation for the measured phenomena. Indeed the pumped area of the cluster is kept constant and the same throughout the measurements. We take care of this fact by exploiting a suitable pumping shape: the disk shaped area centered on the cluster (circular red area in figure \ref{fig 5} (a)) assures that the whole cluster is homogeneously pumped for all the values of $\Theta$.

The fact that the mode shape at $\lambda$ changes with $\Theta$ (Fig. \ref{fig 2}), for a given realization of disorder, can instead be understood by taking into account mode-coupling in open cavities \cite{Conti08, Leonetti_Haus_Gross}: a mode at $\lambda_1$ can excite other modes and force them to oscillate at its own $\lambda_1$, i.e. at a frequency different from their natural resonance. Hence a localized oscillation at $\lambda_1$ spatially broadens subsuming several other modes in other locations. However, lossy (or scarcely pumped) modes are not able to retain energy at frequencies far from their natural resonances (small $\Theta$ regime). On the contrary, when increasing $\Theta$, modes in any point in the structure can sustain oscillations at any frequency and, as they are all coupled, they synchronize, resulting in large scale coherent emission.

This particular mechanism in which a single laser mode alters the oscillation frequencies of its neighbors to extend its effective localization length has never been measured previously, nor has, to the best of our knowledge, the prediction of such behavior been made. The maximum extent achievable by an individual mode is limited by the disordered structure: this is confirmed by the insets of Fig.\ref{fig 3} which compare single mode size at high $\Theta$ with the shape of the corresponding cluster (they turn out to be very similar) and by the greatest achieved value of the inverse participation ratio which is larger for bigger clusters.

\textbf{}

{\it Conclusions ---}
We show evidence of the effect inter mode interaction has on the spatial extent of lasing modes in a RL. The coupling between modes allows the spreading of the energy distribution of a mode into the whole amplifying region. This may be explained by the fact that a stronger mode forces weaker neighbors to oscillate at its resonant frequency, increasing numbers of modes being involved as $\Theta$ increases. The ability to control the onset of a collective oscillation and the mode extent at the micrometer level represents, in our opinion, a road to many potential applications. Future generations of light driven photonic chips, for example, may take advantage of similar phenomena needing strongly localized and weakly interacting resonances when used to store information while a strong inter-mode interaction could be exploited at the elaboration step. Other possible applications for clusters as light sources is in the development of tunable optical coherence tomography techniques, and in speckle-less coherent imaging.

\begin{acknowledgments}
The research leading to these results has received funding from the ERC under the EC's Seventh Framework Program (FP7/2007-2013) grant agreement n.201766, EU FP7 NoE Nanophotonics4Enery Grant No 248855; the Spanish MICINN CSD2007-0046 (Nanolight.es); MAT2009-07841 (GLUSFA) and Comunidad de Madrid S2009/MAT-1756 (PHAMA).
\end{acknowledgments}


%

\end{document}